\documentclass[draftcls, onecolumn, 12pt]{IEEEtran}

\usepackage{bbm}
\usepackage{amssymb}
\usepackage{bbding}
\usepackage{amsfonts}
\usepackage[cmex10]{amsmath}
\usepackage{mathrsfs}
\usepackage{graphicx}
\usepackage{cite}
\usepackage{amsmath}
\usepackage{subfigure}
\usepackage{slashbox}
\usepackage{ccaption}
\usepackage{booktabs}
\usepackage{array}
\usepackage{longtable}
\usepackage{bm}
\usepackage[ruled,linesnumbered]{algorithm2e}

\newtheorem{theorem}{Theorem}
\newtheorem{lemma}{Lemma}

\begin{document}
\baselineskip=24pt

% paper title
% can use linebreaks \\ within to get better formatting as desired
\title{Joint Channel Probing and Proportional Fair Scheduling in Wireless Networks}

% author names and affiliations
% use a multiple column layout for up to three different
% affiliations

%\author{\IEEEauthorblockN{Hui Zhou}
%\IEEEauthorblockA{Dept. of Electronic Eng.\\
%Tsinghua University\\ Beijing, China\\ E-mail:
%h-zhou03@mails.thu.edu.cn} \and \IEEEauthorblockN{ Pingyi Fan}
%\IEEEauthorblockA{Tsinghua University, China;\\
% National Mobile Communications \\Research Laboratory\\
%  Southeast University, China \\
%Email: fpy@tsighua.edu.cn} \\
% \IEEEauthorblockA{K. B. Letaief \\
%Department of ECE \\ HKUST, Hong Kong \\ E-mail: eekhaled@ee.ust.hk}
% }

%\author{\IEEEauthorblockN{Hui Zhou,
% Pingyi Fan}
%\IEEEauthorblockA{Department of Electronic Engineering, Tsinghua
%University, Beijing, China
%\\}
%E-mail:h-zhou03@mails.thu.edu.cn, fpy@tsinghua.edu.cn }

\author{\IEEEauthorblockN{Hui Zhou, Pingyi Fan, Dongning
Guo \footnote{H. Zhou and P. Fan are with the Department of
Electronic Engineering, Tsinghua University, Beijing, 100084, China
(e-mail: h-zhou03@mails.thu.edu.cn; fpy@tsinghua.edu.cn) \\
\indent D. Guo is with the Department of Electrical Engineering and
Computer Science, Northwestern University, Evanston, IL 60208,
U.S.A. (e-mail: dguo@northwestern.edu) }  }

% \\ \IEEEauthorblockA{\IEEEauthorrefmark{1}\IEEEauthorrefmark{2}Department
%of Electronic Engineering, Tsinghua University, Beijing, China
%\\\IEEEauthorrefmark{3}Department of EECS, Northwestern University, Evanston, IL,
%USA} \\ E-mail:\IEEEauthorrefmark{1}h-zhou03@mails.thu.edu.cn,
%\IEEEauthorrefmark{2}fpy@tsinghua.edu.cn,
%\IEEEauthorrefmark{3}dguo@northwestern.edu

}

% conference papers do not typically use \thanks and this command
% is locked out in conference mode. If really needed, such as for
% the acknowledgment of grants, issue a \IEEEoverridecommandlockouts
% after \documentclass

% for over three affiliations, or if they all won't fit within the width
% of the page, use this alternative format:
%
%\author{\IEEEauthorblockN{Michael Shell\IEEEauthorrefmark{1},
%Homer Simpson\IEEEauthorrefmark{2},
%James Kirk\IEEEauthorrefmark{3},
%Montgomery Scott\IEEEauthorrefmark{3} and
%Eldon Tyrell\IEEEauthorrefmark{4}}
%\IEEEauthorblockA{\IEEEauthorrefmark{1}School of Electrical and Computer Engineering\\
%Georgia Institute of Technology,
%Atlanta, Georgia 30332--0250\\ Email: see http://www.michaelshell.org/contact.html}
%\IEEEauthorblockA{\IEEEauthorrefmark{2}Twentieth Century Fox, Springfield, USA\\
%Email: homer@thesimpsons.com}
%\IEEEauthorblockA{\IEEEauthorrefmark{3}Starfleet Academy, San Francisco, California 96678-2391\\
%Telephone: (800) 555--1212, Fax: (888) 555--1212}
%\IEEEauthorblockA{\IEEEauthorrefmark{4}Tyrell Inc., 123 Replicant Street, Los Angeles, California 90210--4321}}

% use for special paper notices
%\IEEEspecialpapernotice{(Invited Paper)}

% make the title area
\maketitle

\begin{abstract}
%\boldmath
The design of a scheduling scheme is crucial for the efficiency and
user-fairness of wireless networks.  Assuming that the quality of
all user channels is available to a central controller, a simple
scheme which maximizes the utility function defined as the sum
logarithm throughput of all users has been shown to guarantee
proportional fairness. However, to acquire the channel quality
information may consume substantial amount of resources. In this
work, it is assumed that probing the quality of each user's channel
takes a fraction of the coherence time, so that the amount of time
for data transmission is reduced. The multiuser diversity gain does
not always increase as the number of users increases. In case the
statistics of the channel quality is available to the controller,
the problem of sequential channel probing for user scheduling is
formulated as an optimal stopping time problem. A joint channel
probing and proportional fair scheduling scheme is developed. This
scheme is extended to the case where the channel statistics are not
available to the controller, in which case a joint learning, probing
and scheduling scheme is designed by studying a generalized bandit
problem. Numerical results demonstrate that the proposed scheduling
schemes can provide significant gain over existing schemes.
%This paper studies the impact of the cost of user rate probing on
%the scheduling. A joint probing and scheduling scheme is proposed,
%which achieves the optimal throughput under the proportional
%fairness constraint. Simulations show that  when probing cost is
%taken into account, the multi-user diversity gain is not always
%increasing as users' size increases. Hence the cost of channel rate
%acquisition should not be ignored in the design and analysis of
%scheduling problems. We also show that the proposed scheme
%significantly outperforms existing schemes which also consider the
%channel probing cost.
\end{abstract}

\IEEEpeerreviewmaketitle

\section{Introduction}
Efficient and fair scheduling is  important for wireless systems
with limited resources and heterogeneous user conditions. A large
class of  resource allocation schemes with fairness considerations
are obtained by maximizing some utility functions of the throughput
\cite{mo}. In particular, proportional fairness is achieved when the
utility is the sum of the logarithm of the users' throughput. In
existing third generation wireless systems, like EV-DO and HSDPA,
proportional fair (PF) scheduling scheme is employed at the base
station to schedule downlink traffic to mobile users. The PF scheme
strikes a good balance between throughput efficiency and fairness by
exploiting the multiuser diversity \cite{viswanath} and the
game-theoretic equilibrium \cite{kelly}. Analysis and applications
on PF scheduling have been extensively explored from various aspects
due to its favorable performance and low implementation complexity.
For example, there have been studies of the convergence and
optimality \cite{kushner2}, stability \cite{brost_sta}, throughput
 \cite{choi} and capacity region
\cite{liu}  of PF scheduling.
%Application related work
%aim to implement the PF algorithm on practical systems, like
%orthogonal frequency-division multiplexing (OFDM) systems,
%multi-channel systems and multi-cell networks.

Most previous work on PF scheduling assume that the instantaneous
channel quality information (CQI) of all users is known to the
scheduler at no cost. In practice, however, acquiring the CQI often
consumes a significant amount of resources in terms of time,
bandwidth and power. It is  important to understand the impact of
the cost when the number of users is large, because the cost may
scale linearly with the user population. The goal of this work is to
answer the following two questions: 1) to what extent will the CQI
acquisition affect the scheduling? and 2) how to probe and schedule
the users to achieve the best performance with proportional
fairness?

There have been related works on the impact of the channel
uncertainty on the communication systems. The loss of throughput
caused by poor estimates of channel quality is quantified in
\cite{chan}.  Joint channel probing and user scheduling  has also
been addressed  recently. Several schemes with the objective of
maximizing the system throughput have been designed in
\cite{guha,chang,chaporkarMaxThr,chen}. And the authors of
\cite{gopalan,ouyang, chaporkar} propose schemes for stabilizing the
queues and characterize the network throughput region.  In contrast
to the preceding works, the goal of this paper is to design a
proportional fair scheduling scheme
 which takes into account the cost of channel probing. Our previous
 work \cite{zhou} has shown the scheme and its performance roughly. In this
 paper, we not only present the derivation of the scheme with
rigorous arguments, but also show its asymptotic behavior and the
optimality with theoretical rigor. In addition, the scheme is
extended to a more generalized scenario. The organization and main
contributions of this work are as follows:
\begin{itemize}
\item Section II describes the network model.
\item  In Section III, we assume the prior distribution of CQI is known
to the scheduler, and formulate the problem of sequentially probing
user channels to make scheduling decision as a stopping time
problem. A simple scheme based on maximizing the sum logarithm
throughput of all users is shown to guarantee proportional fairness
and convergence.  The scheduling gain of the scheme is determined
analytically.  Further reduction of computational complexity is also
discussed.
\item In Section IV, the statistics of the CQI is assumed not to be available to the
scheduler. The problem is formulated as a generalized bandit
problem, and  a joint learning, probing and scheduling scheme is
proposed.
\item In Section V,  significant advantages of the proposed schemes
 are demonstrated using numerical experiments.
In typical scenarios where the statistics of the CQI are not
available, the joint learning, probing and scheduling scheme
achieves almost the same performance as that in the case where the
statistics are known.
\end{itemize}
%
%The rest of the paper is organized as follows. The network model is
%detailed in Section II. Section III considers the scenario where the
%channel statistics is known by the base station. The extension to
%the unknown channel statistics case is developed in Section IV.
%Section V presents numerical results that confirm our theoretical
%claims. Finally, Section VI concludes the paper.

%However, the structural property study of optimal scheme for
%maximizing queue normalized throughput  by Chaporka et
%al\cite{chaporkar} provides us a framework to follow. Their work
%inspires us to concern the exploration and exploitation tradeoff in
%the context of utility maximization.

\section{The Network Model}
Consider a wireless system with one controller and $K$ users with
time-varying channel quality, such as in
 the downlink of a cellular system. Let time be divided into
unit-length slots and only one user can be served in each slot. As
in most related work (e.g., \cite{kushner2} and \cite{choi}),  the
transmit power is assumed to be fixed so that dynamic power
allocation is not considered. Thus the achievable rate is only
determined by the instantaneous channel quality.
 Moreover, we assume saturated traffic for all users.
%Congestion control and queue management
%issues are not considered at all.

Assume slow fading, where the duration of a slot is much shorter
than the channel coherence time, so that the channel quality remains
constant during each slot. We make the following {\em homogeneous
rate assumption} that the rate of each user normalized by its mean
value follows the same distribution:

%In this paper, we employ a relevant assumption that all users'
%achievable rates follow the same type of distribution as follows.

%For user $k$, we assume that $\{R_k(n)\}_{n=1}^{\infty}$ is an
%i.i.d. sequence. The vector of user rates is denoted as $\bm R(n) =
%[R_1(n) \cdots R_K (n)]$. Ignoring the slot index $n$, we use the
%random variable $R_k$ to denote the canonical value of $R_k(n)$. The
%rates $R_1,\dots,R_K$ are independent but not necessarily
%identically distributed. In this paper, we employ a relevant
%assumption that all $R_k$ follow the same type of distribution as
%follows.

\emph{(A1)} Let $X_1,\dots,X_K$ be independent identically
distributed (i.i.d.) non-negative random variables with unit mean
value.  Let $r_1,\dots,r_K\ge0$ be constants.  Let $R_k=r_kX_k$ for
$k=1,\dots,K$.  The achievable rates $\{R_k(n)| k=1,\dots,K;
n=1,2,\dots\}$ are independent.  For every user $k$, the rates over
the time slots, $R_k(1), R_k(2), \dots$, are i.i.d.\ following the
same distribution as that of $R_k$. Clearly, $\mathbb{E} R_k(n) =
r_k$.

%\emph{(A1)} For every $k=1,\dots,K$ and every $n=1,2,\dots$, the
%achievable rate $R_k(n)$ follows the same distribution as $r_k X_k$,
%where $r_k $ is a constant and  $X_1, ..., X_K$ are independent
%identically distributed random variables with unit mean value.

%$R_k = r_k X_k$, where $r_k=\mathbb{E}R_k$, and $\{X_k\}_{k=1}^K$
%are i.i.d. random variables with $\mathbb{E} X_k =1$.
%We use a unit-mean valued random variable $X$ to denote the
%characteristic distribution of all channel rates.
% We assume that the
%transmitter only knows the distribution $F_{R_k}(\cdot) =
%F_{x}(\frac{1}{r_k} \cdot)$ of the channel rates.

%For each user that has been probed, the channel rate $R_k(n)$
%\footnote{If user $k$ is probed, then $R_k(n)$ is no longer a random
%variable, but an observed value. The same symbols are used without
%causing ambiguity.} is available at the transmitter.

The instantaneous achievable rates of all users are not known
\emph{a priori}. During each slot $n$, to obtain the achievable rate
$R_k(n)$ requires the scheduler to probe the channel of user $k$
using a fraction $\beta$ of the slot. Let $I_k(n)$ be an indicator
of the event that user $k$ is scheduled for transmission in slot
$n$. Let $J(n)$ denote the number of probed users in slot $n$.  The
amount of data transmitted to or by user $k$ during slot $n$ is $
B_k(n)=(1-J(n)\beta) R_k(n) I_k(n), $ which is nonzero for only one
user during each slot. The throughput of user $k$ averaged over $n$
slots is thus
\begin{equation}  \label{eqn_Tk_def}
T_k(n) = \frac1n \sum^n_{j=1} B_k(j).
\end{equation}
%Then users' throughput is $\bm T(n) = \frac{1}{n} \sum_{j=1}^n \bm B(j)$.
%Proportional fairness  is achieved by means of maximizing the
%utility function
%\begin{equation}  \label{eqn_PFutility}
%\lim_{n \to \infty} u(\bm T(n)),
%\end{equation}
%where  $u(\bm T(n))= \sum_{k=1}^K \ln(T_k(n))$.

\section{Joint Probing and Scheduling with Known Channel Statistics}
 In this section, we consider the case where
the statistics of $\bm R=[R_1,\dots,R_K]$ is known to the scheduler
and design a  proportional fair scheme.
%With this setting, the
%design of optimal joint probing and scheduling problem can be
%formulated as an optimal stopping time problem.

\subsection{The Algorithm}
\label{sec_jps_formulation_scheme} Consider first a scheme which
maximizes the utility defined as the sum logarithm throughput:
\begin{align} \label{eqn_PFutility}
u(\bm T(n)) = \sum^K_{k=1} \ln T_k(n) \ .
\end{align}
Note that by (\ref{eqn_Tk_def}),
\begin{equation}  \label{eqn_tkn_update}
T_k(n) = \frac{n-1}{n} T_k(n-1) + \frac{1}{n} B_k(n).
\end{equation}
So that the increase of the utility function after the $n$-th slot
is
\begin{align} \nonumber  \label{eqn_utility_increase}
  & u(\bm T(n)) - u(\bm T(n-1)) \nonumber \\
  = & \sum_{k=1}^{K} ( \ln T_k(n) - \ln T_k(n-1)  ) \nonumber \\
 = & \sum_{k=1}^K \ln \left( \frac{n-1}{n} + \frac{1}{n} \frac{B_k(n)}{T_k(n-1)} \right) \nonumber \\
 = & \sum^K_{k=1} \ln \left( \frac{n-1}{n} + \frac{1-\beta J(n)}{n} s_k(n) I_k(n)
 \right),
\end{align}
where the throughput-normalized rate is
\begin{equation} \label{eqn_sn}
s_k(n)=\frac{{R_k(n)}}{T_k(n-1)}.
\end{equation}
 Since the indicator $I_k(n)$ is zero for all but one
user $k$ in each slot, one can see that to greedily maximize the
utility increment at time slot $n$, we should schedule the user with
the maximum $s_k(n)$, which is the classical PF scheduling
algorithm.
%\begin{equation} \label{eqn_QnPF}
%k^* = \arg \max s_k(n).
%\end{equation}

%\begin{align} \nonumber
%  \triangle f & \triangleq f(\bm T(n)) - f(\bm T(n-1)) \nonumber \\
% & =  \nabla f(\bm T(n-1)) \cdot  (\bm T(n) - \bm T(n-1)) + o\left( 1/n
% \right).
%\nonumber
%%\\
%%= & \nabla f(\bm T(n-1)) \cdot \left[ \frac{(n-1) \bm T(n-1) + \bm
%%B(n) }{n} - \bm T(n-1) \right] \nonumber \\
%%= & \nabla f(\bm T(n-1)) \cdot \frac{1}{n} \left[ \bm B(n) - \bm
%%T(n-1) \right] \nonumber \\
%%= & \frac{1}{n} \nabla f(\bm T(n-1)) \cdot \bm B(n) - \frac{K}{n}.
%%\nonumber
%\end{align}
%Noting that $n\bm T(n)-(n-1)\bm T(n-1) = \bm B(n)$, we have
%\begin{align} \nonumber
%\triangle f &= \nabla f(\bm T(n-1)) \cdot \frac{1}{n} (\bm B(n)-\bm
%T(n-1)) + o\left( 1/n  \right)
%\nonumber  \\
%&=\frac{1}{n} \nabla f(\bm T(n-1)) \cdot \bm B(n) - \frac{K}{n} +
%o\left( 1/n \right) . \nonumber
%\end{align}
%Clearly, the optimal scheme $\Gamma(n)$ in slot $n$ must satisfy
%\begin{equation}  \label{eqn_QnPF}
%\Gamma(n) \in \arg \max \nabla f(\bm T(n-1)) \cdot \bm B(n).
%\end{equation}
%
% As only one user can be selected as the
%destination in a slot, the joint probing and scheduling scheme
%basically includes two tasks: to determine the order in which users
%are probed, and to select one user as the destination at a proper
%(stopping) time. Noting that
%\begin{align}  \nonumber
%& \nabla f(\bm T(n-1)) \cdot \bm B(n)
%\nonumber \\
%=& \sum^K_{u=1} \frac{B_k(n)}{T_k(n-1)}
%\nonumber \\
%=& (1-J(n)\beta) \sum\limits_{k=1}^K \frac{ R_k(n) I_k(n)}{T_k(n-1)}
%. \nonumber
%\end{align}
However, due to the assumption that the instantaneous rates $R_k(n)$
are unknown a priori, we can only probe the users rates and obtain
$s_k(n)$ one by one in each slot. We formulate the following optimal
stopping time problem \cite{ferguson}. Note that the scheduling
decision made in one slot has no impact on future realization of the
rates, it suffices to consider one arbitrary slot and omit the time
index $n$. For the scheduler, the joint probing and scheduling
problem at the beginning of the time slot is defined by two objects:

(i) The independent throughput-normalized rates $s_1,\dots,s_K$.

(ii) A sequence of positive-valued reward functions $y_1,\dots,y_K$,
where if $j$ channels have been probed to reveal their
throughput-normalized instantaneous rates $t_1,\dots,t_j$, the
reward of terminating the probing phase and schedule the best user
found so far is
\begin{equation}  \label{eqn_yk}
y_j(t_1,\dots,t_j) = (1-j\beta)\max(t_1,\dots,t_j).
\end{equation}

The theory of optimal stopping is concerned with determining the
stopping time $J$ to maximize the expected reward $\mathbb E
[y_{J}]$. The maximum number of probings in every slot is $J_{max} =
\min(K,\left\lfloor 1/\beta \right\rfloor)$. Compared with the
classical optimal stopping problem, the formulation above is more
general in the sense that the probing order of $s_k$ is not
deterministic.
 Hence the joint probing and
scheduling scheme basically includes two tasks in each slot: to
determine the order in which users are probed, and to select one
user as the destination at a proper (stopping) time. Recalling the
objective of maximizing the expected $y_j$,  the user with the
largest $\mathbb{E} [s_k(n)]$  should be probed first, and then the
second largest and so on. From Assumption (A1), we know $\bar s_k(n)
\triangleq \mathbb{E} [s_k(n)] = r_k / T_k(n-1)$. Hence the probing
order is $\pi(n) =( k_1,\cdots,k_K)$ such that $\bar s_{k_1}(n) \ge
\dots \ge \bar s_{k_K}(n)$.  Now that the probing order has been
determined, the decision on when to stop can be addressed by
investigating the structural property of the problem.

%\begin{definition} (\emph{monotone stopping problem} \cite[Chapter
%5]{ferguson}  ) Let $\mathcal {E}_j$ denote the event
%\begin{equation}  \label{eqn_monoDef}
%\left \{ y_j(s_{k_1},\cdots, s_{k_j} ) \ge \mathbb E [
%y_{j+1}(s_{k_1},\cdots, s_{k_{j+1}} ) | s_{k_1},\cdots, s_{k_j} ]
%\right \}.
%\end{equation}
%The stopping problem is monotone if $\mathcal{E}_j \subseteq
%\mathcal{E}_{j+1}$.
%\end{definition}

\begin{theorem}  \label{theo_monote}
Under the homogeneous rate assumption (A1), the joint probing and
scheduling problem is a monotone stopping problem \cite[Chapter
5]{ferguson}, which means that, if $\mathcal{E}_j$ denotes the event
\begin{equation}  \label{eqn_monoDef}
\left \{ y_j(s_{k_1},\cdots, s_{k_j} ) \ge \mathbb E [
y_{j+1}(s_{k_1},\cdots, s_{k_{j+1}} ) | s_{k_1},\cdots, s_{k_j} ]
\right \},
\end{equation}
 then $\mathcal{E}_j \subseteq \mathcal{E}_{j+1}$ for $0 \le j \le J_{max}-1$.
\end{theorem}

\emph{Proof:} See appendix \ref{app_proof_monote}.

Now the problem has been proved to be monotone, then from the
\cite[Theorem 1, Chapter 5]{ferguson}, the one-state look-ahead rule
is optimal. The one-stage look-ahead  rule is the one that stops if
the reward for stopping at current stage is at least as large as the
expected reward of continuing one stage and then stop.
Mathematically, the  rule is described by the stopping time. Let
$w_j$ denote the largest value of the observed
 throughput-normalized rate after probing $j$ users
 and  $a \vee b \triangleq \max(a,b)$, the optimal stopping time is
\begin{equation}  \label{eqn_stoppingtime2}
J^* = \min \left\{ j\ge 0: (1-j\beta)w_j \ge
(1-(j+1)\beta)\mathbb{E} \left [ w_j \vee
\frac{R_{k_{j+1}}}{T_{k_{j+1}}(n-1)} \bigg | w_j \right ] \right\},
\end{equation}
which solves the stopping problem  almost surely in each slot.
Precisely, the optimal PF joint probing and scheduling (JPS-PF)
scheme is described as Algorithm 1.
\begin{algorithm}
\caption{ JPS-PF} \label{alg_jps_pf} \SetKwRepeat{Repeat}{do}{while}
\textbf{Initialization:}  $T_k(0) \leftarrow 1$  for
$k=1,\cdots,K$\;
 \For{$n=1,2,\cdots$} {
    $\bar s_k(n) \leftarrow r_k / T_k(n-1)$. Sort the throughput-normalized mean rate
           $\bar s_k(n)(k=1,\cdots,K)$ in the descending order:
           $\bar s_{k_1}(n) \ge \dots \ge \bar s_{k_K}(n)$
           \;
    $j \leftarrow 0$, $w \leftarrow 0$ \;
    \Repeat{$(1-j\beta)w < (1-(j+1)\beta)
           \mathbb{E} \left [ w \vee \frac{R_{k_{j+1}}}{T_{k_{j+1}}(n-1)} \right ]$}
    {
         $j \leftarrow j+1$ \;
         Probe  user $k_j$ and get the rate $R_{k_j}(n)$ \;
         $w \leftarrow w \vee R_{k_j}(n) / T_{k_j}(n-1) $ \;
    }
    Transmit to user $k_j$. Update $\bm T(n)$ \;
}
\end{algorithm}

\subsection{On the Optimality of Algorithm 1}

To present the optimality of Algorithm 1, we need to show the
convergence property.
\begin{theorem} \label{theo_convergence}
Assume (A1). Then for any initial condition, the throughput sequence
$\bm T(n)$ generated under Algorithm 1 converges almost surely to
the limit point $\bm T^*$ of the ordinary differential equation
$\dot{\bm T}(t)=\bm h (\bm T(t))$, where $\bm h(\bm T)=-\bm
T+\mathbb{E}[\bm B(n) | \bm T(n-1) = \bm T]$. Moreover, all users'
steady-state throughput are proportional to their mean rate with an
identical ratio $\kappa$,
\begin{equation} \label{eqn_throughput_ratio}
\frac{T_1^*}{r_1} = \frac{T_2^*}{r_2} = \cdots = \frac{T_K^*}{r_K} =
\kappa.
\end{equation}
\end{theorem}

\begin{IEEEproof}
Let $\bm M(n)=\bm B(n) - \mathbb{E}[\bm B(n)|\bm T(n-1)]$. By
(\ref{eqn_tkn_update}), the update of users' throughput can be
organized in the form of stochastic approximation iteration
\cite[Eqn.~2.1.1]{borkar}:
\begin{align}
\bm T(n) = \bm T(n-1) + a(n) [ \bm h(\bm T(n-1)) + \bm M(n) ],
\nonumber
\end{align}
%\begin{eqnarray} \label{eqn_stoapp}
% \bm T(n+1) &=&  \bm T(n) + \frac{1}{n+1} \{-\bm T(n)+ \mathbb{E}[\bm
% B(n+1) |\bm T(n) ]
%  {} \nonumber\\ {}
%& & + \text{   } \bm B(n+1) - \mathbb{E}[\bm B(n+1)|\bm T(n)] \}
%  {} \nonumber\\ {}
%&=& \bm T(n) + a(n) [ \bm h(\bm T(n)) + \bm M(n+1) ]. \nonumber
%\end{eqnarray}
where $a(n)=1/n$. The  equation above is a standard stochastic
approximation expression. It is easy to verify that $\bm h(\cdot)$
is Lipshitz, the stepsize satisfies $\sum_n a(n)=\infty, \sum_n
a(n)^2 < \infty$ and $\bm T(n)$ is bounded. Furthermore, it is easy
to verify that $\mathbb{E} [ \bm M(n) | \bm M(1),\cdots, \bm M(n-1)]
= 0$, so $\bm M(n)$ is a martingale difference sequence. Now the
throughput update under the proposed scheme satisfies the
assumptions (A1)-(A4) in \cite[Section 2.1]{borkar}, then applying
Theorem 2 in \cite[Section 2.1]{borkar} directly, the convergence
conclusion holds.

Now the convergence of the throughput sequence has been obtained.
The remainder of the proof is by contradiction. Suppose
(\ref{eqn_throughput_ratio}) does not hold at steady state and that
${T_1^*}/{r_1} < {T_2^*}/{r_2} $ without loss of generality.
Consider the throughput path starting at slot $n_0$ which is at
steady state. At this
 time,  $\bar s_l = r_l / T_l^*(l=1,2)$ and $\bar s_1 > \bar s_2$.
 Thus user $1$ is probed first in each slot. From assumption (A1) we know
 that $s_1$ and $s_2$ are of the same type of distribution, but $s_1$
 has a larger mean value. Thus user $1$ is selected for transmission more
 often than user $2$, which would further imply ${T_1 (n_0+n_1) }/{r_1}
  > {T_2(n_0+n_1)}/{r_2} $ after a sufficiently large number ($n_1$) of slots,
 which contradicts the steady state assumption with $T_1^*/r_1 < T_2^*/r_2$.
\end{IEEEproof}

Note that the constant proportionality factor  $\kappa$ is a bridge
connecting the steady-state throughput and the mean-rate. After
obtaining $\kappa$, it is straightforward to evaluate the throughput
and utility. On the other hand, due to the fact that $\kappa$ is a
constant, we have the following corollary from the proof of Theorem
\ref{theo_convergence}.

\newtheorem{corollary}{Corollary}
\begin{corollary} \label{theo_IdenticalSelectedProb}
Under Algorithm 1, the probability that each user is selected as the
destination is identical as $1/K$.
\end{corollary}

%This result looks kind of counter intuitive, because from the
%description of Algorithm 1, it seems that the user which are probed
%first has a larger probability of being selected as that
%destination. However, due to the fact that the user probing order is
%decided dynamically by the instantaneous-throughput-normalized mean
%rate $s_n$, as given in (\ref{eqn_sn}), the user with a smaller
%$T_k(n-1)$ has a larger ${r_k}/{T_k(n-1)}$, meaning that this user
%will be probed and picked with higher priority, which will increases
%this user's throughput in return. When the algorithm runs for a
%sufficiently long time, the whole system achieves the steady state
%where all  ${r_k}/{T_k(n-1)}$ are identical and hence the
%probabilities of being selected are identical for all users, too.

Algorithm 1 is asymptotically optimal in the following sense:
\begin{theorem} \label{theo_optimality}
Assume (A1). Then $\bm T^*$ maximizes the PF utility $u(\cdot)$ over
the rate region generated by all joint probing and scheduling
schemes.
\end{theorem}

\begin{IEEEproof}
Let $\mathcal{S}$ denote the set composed of all the feasible
schemes $\Gamma$ under the assumption that only one user can be
selected in one slot. The developed scheme in this paper is denoted
as $\Gamma^*$. We have shown in the derivation of Algorithm 1 that
$\Gamma^*$ is optimal for solving the monotone stopping problem in
each slot, that is, it maximizes $B_k(n)/T_k(n-1)$ in slot $n$
almost surely. Due to the constraint that only one user can be
scheduled in one slot, we can see that the developed scheme
$\Gamma^*$ satisfies
\begin{equation}  \label{eqn_gammastar}
\Gamma^* \in  \arg \max_{\Gamma \in \mathcal{S} } \sum_{k=1}^K
\frac{B_k^{(\Gamma)}(n)}{T_k(n-1)},
\end{equation}
where $B_k^{(\Gamma)}(n)$ is  the number of bits transmitted to user
$k$ in slot $n$ under the scheme $\Gamma$. Recalling the definition
of the utility function in (\ref{eqn_PFutility}), it can be found
that
\begin{equation}  \label{eqn_gradient}
\sum_{k=1}^K \frac{B_k^{(\Gamma)}(n)}{T_k(n-1)} = \nabla u(\bm
T(n-1)) \cdot \bm B^{(\Gamma)}(n),
\end{equation}
which means that the scheme chooses a decision maximizing the scalar
product of $\bm B^{(\Gamma)}(n)$ and the gradient $\nabla u(\bm
T(n-1))$.

The \emph{gradient scheduling} algorithm developed by Stolyar
\cite{stolyar} is that,  at time $n$ the controller chooses a
decision $\Gamma(n) \in \arg \max\limits_{\Gamma} \nabla u(\bm
T(n-1)) \cdot \bm B ^{(\Gamma)}(n)$. Let $\tilde{\bm T}$ denote the
solution to the problem
\begin{align}
\max & \text{~~~~}u(\bm T) \nonumber \\
s.t. & \text{~~~~} \bm T \in \mathcal{V} \nonumber,
\end{align}
where $\mathcal{V}$ is the system rate region, i.e., the set of all
feasible long-term service rate vectors.  Then the \cite[Theorem
2]{stolyar} shows that the expected average service rates under the
gradient scheduling algorithm converges in probability to
$\tilde{\bm T}$.

By (\ref{eqn_gammastar}) and (\ref{eqn_gradient}), one can see that
the joint probing and scheduling algorithm in this paper belongs to
the gradient scheduling algorithm. From the convergence of Algorithm
1, we know $\bm T^* = \tilde{\bm T}$. Then the achieved throughput
$\bm T^*$ maximizes the PF utility function asymptotically.
\end{IEEEproof}

\subsection{A Static Threshold Criteria} \label{sec_static_scheme}

Note that in Algorithm 1, after each probe, the scheduler needs to
evaluate the expectation in (\ref{eqn_stoppingtime2}) which depends
on the channel realizations. Further reduction in the computational
complexity is possible by simply comparing the highest normalized
rate against a sequence of deterministic thresholds, in lieu of
computing (\ref{eqn_stoppingtime2}).  Consider the steady-state case
where users' throughput is exactly $\bm T^*$. Note that by Theorem
\ref{theo_convergence},
\begin{align}  \label{eqn_Zu_Def}  \nonumber
 \frac{R_{k_{j+1}}}{T_{k_{j+1}}(n-1)}  =
\frac{R_{k_{j+1}}}{T_{k_{j+1}}^*}  , \nonumber
\end{align}
%\begin{equation}  \label{eqn_Zu_Def}
%\frac{R_k}{T_k^*}=\frac{R_k}{\kappa r_k}=\frac{r_k X_k}{\kappa
%r_k}=\frac{1}{\kappa} X_k \triangleq Z_k.
%\end{equation}
%From the definition of $X_k$, we know that $\{Z_k\}_{u=1,\cdots,K}$
%are i.i.d. with mean $1/ \kappa$ and distribution $F_X(\kappa z)$.
%The physical interpretation of $Z_k$ is that it is a random variable
%characterizing the statistical properties of the throughput-normalized
%rate in the steady state.
%Then in the steady state, the term
%$\frac{R_{u^{(k+1)}}}{T_{u^{(k+1)}}(n-1)} $ in Eqn.
%(\ref{eqn_proborTrans}) becomes $
%\frac{R_{u^{(k+1)}}}{T_{u^{(k+1)}}^*} = Z_{u^{(k+1)}} = Z_1$, which
%is independent of user index.
which is identically distributed as $X_1/\kappa$. For $0 \le j \le
J_{max}-1$, the inequality of $w_j$ in (\ref{eqn_stoppingtime2})
reduces to
\begin{equation} \label{eqn_staticDetermine}
(1-j\beta)w_j \ge (1-(j+1)\beta)\mathbb{E}[ \max( w_j,
\kappa^{-1}X_1) | w_j ].
\end{equation}

\begin{figure}[h]
\centering
\includegraphics[width=4.7in]{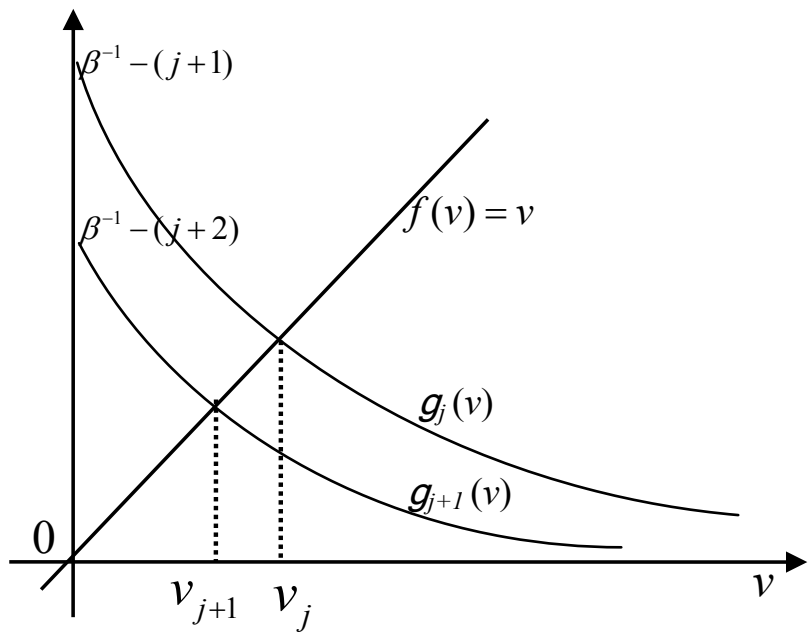}
\caption{Illustration of the solution to inequality
(\ref{eqn_Y_inequality}).} \label{fig_Y_solution} \vspace{0pt}
\end{figure}

It turns out that (\ref{eqn_staticDetermine}) can be reduced to
comparing $\kappa w_j$ with a static threshold $v_j$, which can be
determined as follows. Let $F_X(\cdot)$ denote the cumulative
distribution function (CDF) of $X_k$. Then
%\begin{eqnarray}
%\mathbb{E}\left [ \max \left(w, \frac{X}{\kappa} \right) \right ] &
%= & w \text{Pr}
%\{Z_1 \le w\} + \int_w^{\infty} (z-w) dF_X(\kappa z) {} \nonumber\\
%{} &=& w+ \int_w^{\infty}(z-w) dF_X(\kappa z). \label{eqn_maxWZ}
%\end{eqnarray}
\begin{align}
\mathbb{E}\left [ \max \left(w_j, \frac{X_1}{\kappa} \right) \bigg |
w_j \right ] = w_j+ \int_{\kappa w_j}^{\infty} \left
(\frac{x}{\kappa}-w_j \right ) dF_X(x). \label{eqn_maxWZ}
\end{align}
So that (\ref{eqn_staticDetermine}) can be rewritten as
\begin{equation}  \label{eqn_W_inequality}
 (1-j\beta)w_j \ge (1-(j+1)\beta) \left[ w_j + \int_{\kappa w_j}^{\infty} \left (\frac{x}{\kappa}-w_j \right )
dF_X(x) \right],
\end{equation}
%\begin{equation} \label{eqn_W_inequality2}
%\Rightarrow w \ge \left[ {\beta}^{-1} - (k+1) \right]
%\int_w^{\infty}(z-w) dF_X(\kappa z).
%\end{equation}
or, equivalently,
\begin{equation} \label{eqn_Y_inequality}
\kappa w_j \ge g_j(\kappa w_j),
\end{equation}
where
\begin{equation}  \label{eqn_gjv}
g_j(v)=\left[ {\beta}^{-1} - (j+1) \right]
\int_v^{\infty}(x-v) dF_X(x).
\end{equation}
It is not hard to check that: (i) $g_j(v)>0$ for $v \ge 0$; (ii)
$g_j(v)$ is a strictly decreasing function of $v$; (iii) $\lim_{v
\rightarrow \infty} g_j(v) = 0$. Then inequality
(\ref{eqn_Y_inequality}) is equivalent to $\kappa w_j \ge v_j$,
where $v_j$ is the cross point of function $f(v)=v$ and $g_j (v)$.
Also, we have $g_j(v) > g_{j+1}(v)$. Then it is easy to verify that
$v_{j+1} < v_j$. The solution to (\ref{eqn_Y_inequality}) is
illustrated in Fig. \ref{fig_Y_solution}.

By observing the structure of (\ref{eqn_gjv}), it is worth pointing
out that the cross point $v_j$ is only determined by $j$, $\beta$
and the CDF $F_X(\cdot)$, i.e., the unit mean valued random variable
$X_j$. And the value of $v_j$ is independent of
 the number of users $K$, the mean rates of all users $r_k$ as
well as the achieved throughput to mean-rate ratio $\kappa$. Hence
if the transmitter knows the distribution $F_X(\cdot)$ , it can
compute $v_j$ in advance.

Now inequality (\ref{eqn_staticDetermine}) can be expressed as $w_j
\ge \frac{1}{\kappa} v_j $ for $0 \le j \le J_{max}-1$, which is
also equivalent to the inequality in (\ref{eqn_stoppingtime2}) in
the steady-state case. Thus the decision on whether to keep probing
or to start transmitting is decided by a static threshold criteria.
For completeness, let $v_{J_{max}}=0$ in order to make sure  the
probing can always be terminated in each slot.  We get the following
static threshold based probing criteria, which can replace the line
9 in Algorithm 1.

\emph{Criteria 1:} After probing $j$ users, if the current value of
the largest normalized rate $w_j \ge \frac{1}{\kappa} v_j$, then the
transmitter transmits to the user with the largest normalized rate;
otherwise it probes the $(j+1)$st user.

 In practice, the scheduler can calculate $v_j$
 in advance but $\kappa$ is unavailable at the
 beginning. One way to estimate $\kappa$ is to start the joint probing
and scheduling using the dynamic criteria in line 9 of Algorithm 1.
After a period of time, the throughput approaches to its
steady-state value. Then the throughput to mean-rate ratio $\kappa$
is obtained and the static threshold criteria can be used
thereafter. Alternatively, $\kappa$ can be determined theoretically
as discussed in the next subsection.

\subsection{The Scheduling Gain} In this section we  analyze the
performance of the proposed scheme theoretically. We define the
\emph{scheduling gain} as the ratio of the achieved throughput to
that using round robin scheduling without probing, which reflects
how much multiuser diversity benefits can be exploited.  The
scheduling gain of the proposed joint probing and scheduling scheme
is $ \frac{T_k^*}{K^{-1}r_k}=\kappa K.$ For a random variable $X$,
let us denote the truncation of $X$ over $[a,b]$ as $[X]_a^b$. Note
that $ \mathbb{E}[X|a \le X \le b] = \mathbb{E}{ [X]_a^b }$.

% The PDF of $[X]_a^b$ is  as $f_{[X]_a^b} = \frac{f(x)I_{x
%\in [a,b]}}{F(b)-F(a)}$

\begin{theorem} \label{theo_kappa}
Under the homogeneous rate assumption (A1),  the scheduling gain of
Algorithm 1 is
\begin{align} \label{eqn_kappa}
\kappa K &=& \sum_{j=1}^{J_{max}} \left[(F_X (v_{j-1} ))^{j - 1} -
(F_X (v_j ))^j \right] (1-j\beta)
 \mathbb{E}  \left\{  \left[ \max \left( [X_1]_0^{v_{j-1}}, \cdots ,
 [X_{j-1}]_0^{ v_{j-1} }, X_j \right)
 \right]_{ v_j }^{\infty} \right\}, \nonumber
\end{align}
%\begin{align} \label{eqn_kappa}
%\gamma^{JPS-PF}  &=& \sum_{k=1}^{J_{max}} \left[(F_X (y_k ))^{k - 1}
%- (F_X (y_{k+1} ))^k \right] (1-k\beta)
%  {} \nonumber\\ {}
%& & \mathbb{E}  \left\{  \left[ \max \left( [X_1]_0^{y_k}, \cdots ,
% [X_{k-1}]_0^{ y_k }, X_k \right)
% \right]_{ y_{k+1} }^{\infty} \right\}, \nonumber
%\end{align}
where $v_j$ is the solution of $v=g_j(v)$.
\end{theorem}

Recall that $J^*$ is the optimal stopping time, that is, the number
of users probed before a user is scheduled. We prove  Theorem
\ref{theo_kappa} using the following supporting lemma.

\begin{lemma} \label{theo_ProProbeSteps}
Using Algorithm 1, the steady-state probability of the event that
$j$ users are probed until transmission is given by
\begin{equation}  \label{eqn_ProPorbeStep}
p_j  = {(F_X (v_{j-1} ))^{j - 1}  - (F_X (v_j ))^j } \text{,  } {1
\le j \leqslant J_{max}} .
\end{equation}
%\begin{equation}  \label{eqn_ProPorbeStep}
%p_k  = \left\{ {\begin{array}{*{20}c}
%    {(F_X (y_k ))^{k - 1}  - (F_X (y_{k+1} ))^k ,} & {1 \le k \leqslant K - 1}  \\
%    {(F_X (y_k ))^k, } & {k = K}  \\
% \end{array} } \right.
%\end{equation}
\end{lemma}

\begin{IEEEproof}
At steady state, all users' throughput-normalized mean rates
${r_k}/{T_k^*}$ are essentially identical. Let $q_j=\text{Pr} \{ J^*
\ge j \}$, i.e., the probability that at least $j$ users are probed
before transmission. Then $q_1=1$. And from Criteria 1, we have for
$j \ge 2$,
%Noting that $\{\kappa^{-1} X_k\}_{k=1}^{K}$ are i.i.d. with
%distribution $F_X(\kappa z)$, we have for $k=2,\cdots,J_{max}$,
%After probing $k$ users, the decision on whether \emph{to probe the
%next} or \emph{to transmit to one of the probed} depends on the
%maximum value of the probed users' normalized rates $R_k /T^*_k =
%\kappa^{-1} X_k$.
\begin{align}
q_j &= \text{Pr}\{ \max(X_1,\cdots, X_{j-1}) <  v_{j-1} \} {}
\nonumber\\ {} &= \text{Pr}\{ X_1 <
 v_{j-1} \} \cdots \text{Pr}\{ X_{j-1} <  v_{j-1} \}
{} \nonumber\\ {}  &= (F_X(v_{j-1}))^{j-1} \nonumber.
\end{align}
Like $v_j$, $q_j$ is also completely determined by the rate
distribution. Clearly, $p_j = q_j-q_{j + 1}$ for $j \le J_{max}-1$
and $p_{J_{max}} = q_{J_{max}}$.
\end{IEEEproof}

\emph{Proof of Theorem \ref{theo_kappa}:} Consider a specific user
$k$. In the steady state, $\dot{\bm T}(t)=0$. Then from Theorem
\ref{theo_convergence}, user $k$'s throughput is given by $T_k^* =
\mathbb{E} [B_k(n)| \bm T^* ]$. Throughout, let $K^*$ denote index
of the user that is selected as destination. Then event $\{K^*=k\}$,
i.e., user $k$ is selected as destination, can be decomposed into
$J_{max}$ exclusive sub events: $\{K^*=k\}
=\bigcup\limits_{j=1,\cdots,J_{max}} \{K^*=k, J^*=j\}$. Then we have
\begin{align}\label{eqn_Tu_Derivation}
T_k^* =& \mathbb{E} [B_k(n)| \bm T^* ] \stackrel{ }{=} \mathbb{E}
[(1-J^*\beta)R_k I_k]
  {} \nonumber\\ {}
\stackrel{ }{=}& \text{Pr} \{ K^*=k \} \mathbb{E} [(1-J^*\beta)R_k |
K^*=k ]
  {} \nonumber\\ {}
\stackrel{(a)}{=}& \frac{1}{K} \mathbb{E} [(1-J^*\beta)R_k | K^*=k ]
  {} \nonumber\\ {}
\stackrel{(b)}{=}& \frac{1}{K} \sum_{j=1}^{J_{max}} \text{Pr} \{
J^*=j \} \mathbb{E} [(1-j\beta)R_k | K^*=k, J^*=j]
  {} \nonumber\\ {}
\stackrel{ }{=}&  \frac{T_k^*}{K} \sum_{j=1}^{J_{max}} p_j
(1-j\beta) \mathbb{E} \left [\frac{R_k}{T_k^*} \bigg | K^*=k, J^*=j
\right ] \nonumber
 {} \nonumber\\ {}
\stackrel{(c)}{=}& \frac{T_k^*}{K} \sum_{j=1}^{J_{max}} p_j
(1-j\beta)
 % {} \nonumber\\ {} &
 \mathbb{E} \left\{ \left[ \max \left( \left[ \frac{ R_1}{T_1^*}
\right]_0^{\frac{v_{j-1}}{\kappa} }, \cdots ,
 \left[ \frac{ R_{j-1}}{T_{j-1}^*}
\right]_0^{\frac{v_{j-1}}{\kappa} }, \frac{ R_j}{T_j^*} \right)
 \right]_{\frac{v_j}{\kappa}}^{\infty} \right\}
  {} \nonumber\\ {}
 \stackrel{(d)}{=}& \frac{T_k^*}{K} \sum_{j=1}^{J_{max}} p_j
(1-j\beta)
 % {} \nonumber\\ {} &
 \mathbb{E} \left\{ \left[ \max \left( \left[ \frac{ X_1}{\kappa}
\right]_0^{\frac{v_{j-1}}{\kappa} }, \cdots ,
 \left[ \frac{ X_{j-1}}{\kappa}
\right]_0^{\frac{v_{j-1}}{\kappa} }, \frac{ X_j}{\kappa} \right)
 \right]_{\frac{v_j}{\kappa}}^{\infty} \right\}
  {} \nonumber\\ {}
\stackrel{(e)}{=}& \frac{T_k^*}{\kappa K} \sum_{j=1}^{J_{max}} p_j
(1-j\beta)
 % {} \nonumber\\ {} &
  \mathbb{E}  \left\{  \left[ \max \left( [X_1]_0^{v_{j-1}}, \cdots ,
 [X_{j-1}]_0^{ v_{j-1} }, X_j \right)
 \right]_{ v_j }^{\infty} \right\} \nonumber,
\end{align}
where (a) follows from Corollary \ref{theo_IdenticalSelectedProb},
(b) from the law of total probability,  (c) from the static
threshold criteria, that is,
 $\{K^*=k, J^*=j\}$ means
that: i) user $k$ has the largest throughput-normalized rate among
the first $j$ users; ii) the first $j-1$ users'
throughput-normalized rates are smaller than $\kappa^{-1}v_{j-1}$
and iii) the largest value of the first $j$ users'
throughput-normalized rates is larger than $\kappa^{-1}v_{j}$, (d)
from $R_k = r_k X_k$ and (\ref{eqn_throughput_ratio}), and (e)
 from the distribution of $X_j$. By replacing $p_j$ with
 (\ref{eqn_ProPorbeStep}) and removing $T_k^*$ from both sides,
  the conclusion of Theorem \ref{theo_kappa} holds.
  {\hspace*{\fill}$\blacksquare$\par}

\section{Joint Learning, Probing and Scheduling  }
Consider the case where the scheduler does not know \emph{a priori}
the statistics of the quality  of  the downlink channels, and thus
has to rely on the history of the probed CQI to decide on the user
probing order and user selection. Under this assumption, the problem
of maximizing the PF utility function is a generalization of the
classical multiarmed bandit problem \cite{berry}.  The problem is a
generalization because in the classical bandit problem, the decision
maker has to decide which of $K$ random process to observe in a
sequential of trials so as to maximize the reward, where the
`observing' operation is equivalent to the `utilizing' operation.
However, in our model, in each slot, the scheduler may probe
(observe) more than one channels (random processes) and then choose
only one for transmission (utilization). The observation does not
always lead to a utilization.

%\textbf{Joint Learning, Probing and Scheduling Problem}.
At the beginning of  slot $n$, i.e., the end of slot $n-1$, let
$M_k(n-1)$ denote the number of time slots in which the channel to
user $k$ has been probed,  and $\mathcal{R}_k(n-1) = \{ R_k^{(1)},
\cdots, R_k^{(M_k(n-1))} \}$ record all the probed samples of the
channel rate of user $k$. Clearly, the cardinality $|
\mathcal{R}_k(n-1) | = M_k(n-1)$. The scheduler keeps updating the
$K$ sets $[ \mathcal{R}_1(n),\cdots, \mathcal{R}_K(n) ]$ from slot
to slot. Also, the scheduler knows the throughput $\bm T(n-1)$ till
the previous slot. The objective is still to find a scheme that
solves the stopping problem in each slot. As analyzed in Section
\ref{sec_jps_formulation_scheme}, there still exists the same two
tasks to find the optimal scheme: determining the user probing order
and selecting one user for transmission. Hence the problem
formulation and scheme design is similar to those in Section
\ref{sec_jps_formulation_scheme}. The only difference is that the
scheduler just has  the sampled values of all channels' rates
instead of the explicit knowledge of the distribution of
$R_k,(k=1,\cdots,K)$, which means that we cannot calculate the
expectations related to $R_k$ directly.  Alternatively, we can only
evaluate the empirical average using the acquired samples of $R_k$,
which readily leads to the index-based policy solution in the
framework of bandit problem.

The index policy, consisting of choosing at any time the stochastic
process with the currently highest  index, is the solution to a
class of bandit problems. Here to find the optimal scheme, we adopt
the similar methodology as in the development of the index-based
policy by Agrawal in \cite{agrawal}. For the decision on the user
probing order, we use the current average reward, i.e., the
throughput-normalized average rate as the index. For the decision on
when to start transmission, we adopt the actually served bits in
current slot, i.e., the product of $1-j\beta$ and the conditional
throughput-normalized-average rate.
 For the convenience of presenting the
algorithm, we define the following two empirical averages
\begin{equation}
\tilde{s}_k(n) \triangleq \frac{1}{M_k(n-1)} \sum_{m=1}^{M_k(n-1)}
\frac{R_k^{(m)}}{T_k(n-1)},
\end{equation}
\begin{equation}
\tilde{e}_k(n,w) \triangleq \frac{1}{M_k(n-1)} \sum_{m=1}^{M_k(n-1)}
\left[ w \vee \frac{R_k^{(m)}}{T_k(n-1)} \right].
\end{equation}
The $\tilde{s}_k(n)$ is used to replace the $\bar{s}_k(n)$ in
Algorithm \ref{alg_jps_pf} and the $\tilde{e}_k(n,w) $ is for
$\mathbb{E} \left [ w \vee \frac{R_{k}}{T_{k}(n-1)} \right ]$ in
Algorithm \ref{alg_jps_pf}. Then a joint PF learning, probing and
scheduling   (JLPS-PF) algorithm is described in Algorithm 2.

\begin{algorithm}
\caption{ JLPS-PF} \label{alg_jlps_pf}
\SetKwRepeat{Repeat}{do}{while} \textbf{Initialization:} $n
\leftarrow \lceil \beta K\rceil.$ For $k=1,\cdots,K$, $T_k(n)
\leftarrow 1$. In the first $n$ slots, sequentially probe each
channel once, making sure that each one of the sets
$\mathcal{R}_k(n), (k=1,\cdots,K)$ is not empty. $M_k(n) \leftarrow
1$ \; \For{$n=\lceil \beta K\rceil + 1,\lceil \beta K\rceil +
2,\cdots$} {
    $\tilde{s}_k(n) \leftarrow \frac{1}{M_k(n-1)} \sum\limits_{m=1}^{M_k(n-1)} {R_k^{(m)}}/{T_k(n-1)}$.
           Sort
           $\tilde{s}_k(n)(k=1,\cdots,K)$ in the descending order:
           $\tilde{s}_{k_1}(n)  \ge \dots \ge \tilde{s}_{k_K}(n) $ \;
    $j \leftarrow 0$, $w \leftarrow 0$ \;
    \Repeat{$(1-j\beta)w < (1-(j+1)\beta) \tilde{e}_{k_{j+1}}(n,w) $}
    {
        $ j \leftarrow j+1$ \;
        Probe  user $k_j$ and get the rate $R_{k_j}(n)$ \;
        $w \leftarrow w \vee R_{k_j}(n) / T_{k_j}(n-1) $ \;
        $\tilde{e}_{k_{j+1}}(n,w) \leftarrow \frac{1}{M_{k_{j+1}}(n)} \sum\limits_{m=1}^{M_{k_{j+1}}(n)}
                \left[ w \vee \frac{R_{k_{j+1}}^{(m)}}{T_{k_{j+1}}(n-1)} \right]$ \;
        $ \mathcal{R}_{k_j}(n) \leftarrow \mathcal{R}_{k_j}(n-1) \cup \{ R_{k_j}(n) \}$,
               $M_{k_j}(n) \leftarrow M_{k_j}(n-1)+1$ \;
    }
    Transmit to user $k_j$. Update $\bm T(n)$ \;
    For $k=k_j+1,\cdots,k_K$, $ \mathcal{R}_{k}(n) \leftarrow \mathcal{R}_{k}(n-1) $,
               $M_{k}(n) \leftarrow M_{k}(n-1)$ \;
}
\end{algorithm}

From the description of Algorithm 2, one may wonder such a
phenomenon may exist that if one user is probed with relatively high
values in the first few slots, then it will have low priority of
being probed afterwards, resulting that the ensemble average of this
channel is always higher than its statistical expectation. However,
this does not happen thanks to the structure of the algorithm
derived from the objective of maximizing the PF utility. As a matter
of fact, if user $k$ is probed and selected less frequently compared
to other users, the achieved throughput $T_k(n)$ will  become small,
which will in return increase its priority of being probed and
selected. In fact, the metric of throughput-normalized rate used in
PF scheduling is a well-balanced rule that guarantees each user is
sampled with sufficiently many times and identical frequencies.
Hence after the Algorithm 2 runs a a sufficiently long time, the
sampled data of each user's channel rate can characterize the
statistics of $\bm R$ well. Then from the law of large number, the
ensemble average converges to the statistical expectation. And the
performance of Algorithm 2 is almost the same as that of Algorithm
1.

\section{Numerical Results}
In this section, we provide some numerical experiments illustrating
the theoretical findings of the previous sections. Our objectives
here are (i) to evaluate
 the performance of the developed schemes with and without channel
statistics; (ii) to  compare the developed  scheme for achieving PF
with some ideal and practical schemes and to quantify the impact of
the cost of CQI on the scheduling. We consider the scenario where
users' rates obey the exponential distributions with average equal
to the user index. The exponential rate assumption is an appropriate
approximation of the Shannon capacity under Rayleigh fading channels
in low SNR regime.

\subsection{Evaluation of the Proposed Algorithms}
Consider $K=20$ users and let the fraction of one probe be
$\beta=0.1$. Up to $J_{max}=10$ users can be probed in each slot.

\begin{figure}[h]
\centering
\includegraphics[width=5.5in]{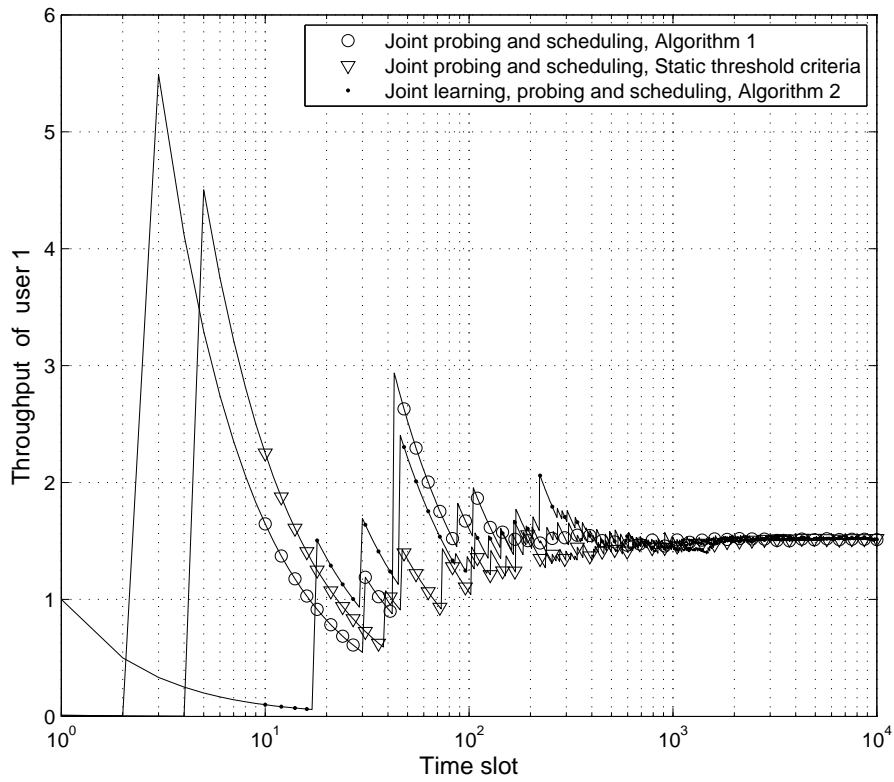}
\caption{The throughput trajectory of user 1 when scheduled with
 Algorithm 1, the static threshold criteria and Algorithm 2
respectively. $N_{slot}=10,000, K=20, \beta=0.1$.}
\label{fig_thrtStaticThresholdLearn}
\end{figure}

Fig. \ref{fig_thrtStaticThresholdLearn} presents a sample throughput
trajectory of user 1 when scheduled with Algorithm 1, the static
threshold criteria given in \emph{criteria 1} and Algorithm 2. The
simulation runs for $10,000$ slots in this experiment. The time axis
is in logarithmic scale to highlight the transient behavior. We can
see that the static threshold criteria works well. The variation of
the throughput diminishes over time as more and more time slots are
included in the averaging. It is worth noting that the low
complexity of the static threshold criteria for solving the optimal
stopping problem comes from the explicit knowledge of the channel
statistics. If this information is not known, or if the distribution
of the channel rate varies over time, we can only adopt the dynamic
criteria given in Algorithm 1.

\begin{figure}[h]
\centering
\includegraphics[width=4.5in]{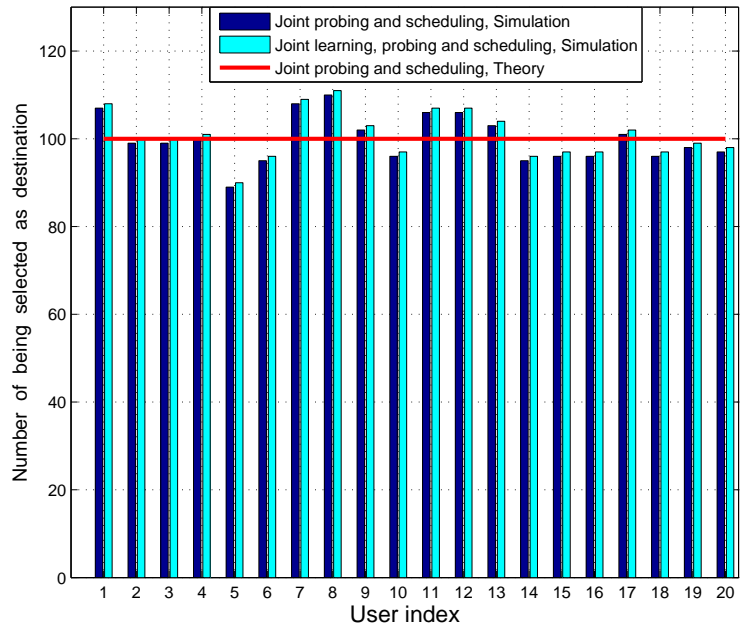}
\caption{The number of slots in which each user is selected as the
destination.
         $N_{slot}=2000, K=20, \beta=0.1$.}
\label{fig_selectNumforUser}
\end{figure}

Fig. \ref{fig_selectNumforUser} illustrates the frequency of each
user being scheduled in a relatively short period of 2000 slots.
Each of the 20 user is selected as the destination for roughly 100
slots. That is, the scheme is fair to all users even within a small
application time window.

\begin{figure}[h]
\centering
\includegraphics[width=4.5in]{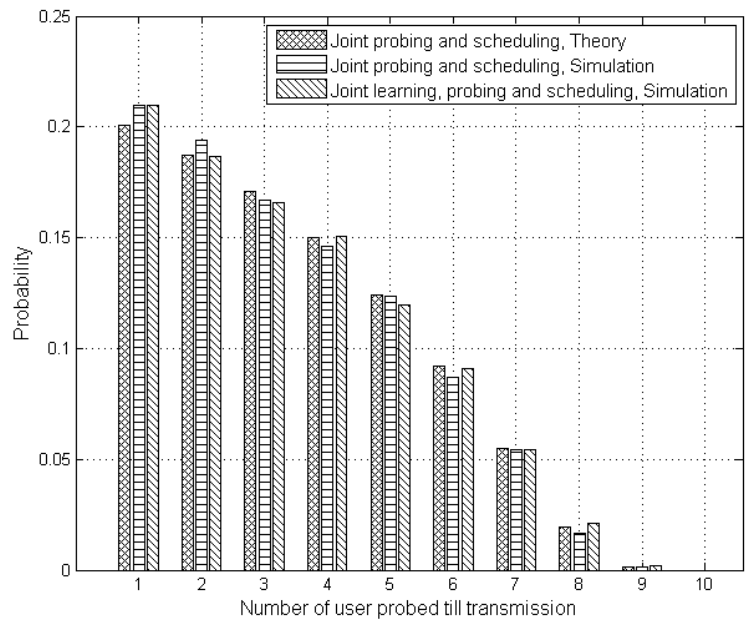}
\caption{The probability that $k$ users have been probed until
transmission.
         $K=20, \beta=0.1$.}
\label{fig_probUserprobe}
\end{figure}

Fig. \ref{fig_probUserprobe} presents the probability that $k$ users
have been probed until transmission. The theoretical results are
from Lemma \ref{theo_ProProbeSteps}. The figure shows that both the
Algorithm 1 and Algorithm 2 coincide with the theoretical results.
We observe from the figure that the probability decreases sharply as
the probing step approaches $J_{max}$.

\begin{figure}[h]
\centering
\includegraphics[width=4.7in]{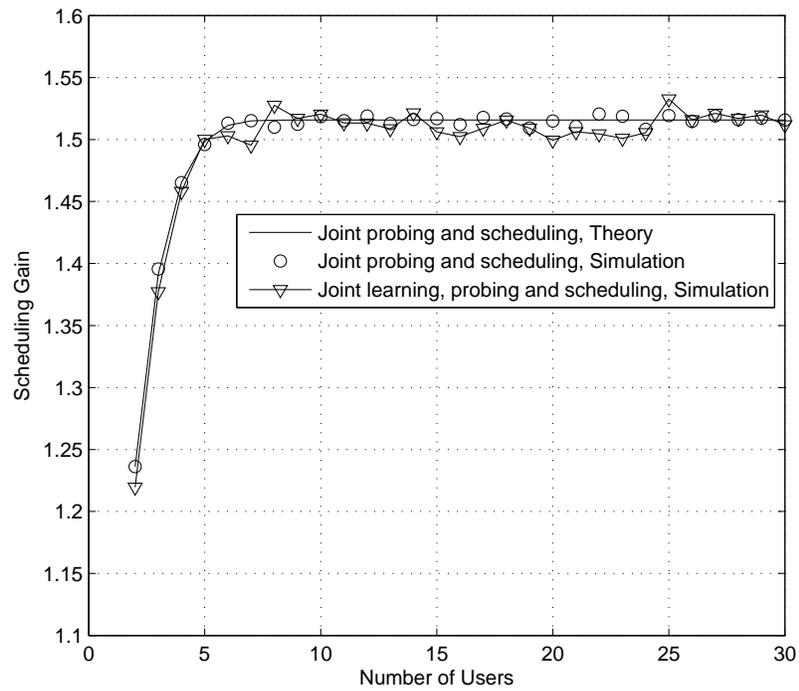}
\caption{The scheduling gain comparison between Algorithm 1,
Algorithm 2 and
 theoretical results. $\beta=0.1$.}
\label{fig_gainLearn}
\end{figure}

Fig. \ref{fig_gainLearn} plots the scheduling gain of the proposed
algorithms versus the number of users in the system. The simulation
runs for 20,000 slots. In fact the simulation result matches the
analytical result of Theorem 4 quite well.  Also, we note the
scheduling gain remains about the same for more than 9 users.
Because at this time, the cost of user probing is dominant and the
scheme always tries to carry out the user probing till the end.

%From all the simulations in this subsection, we can see that the analytical
%results can reflect the real simulations using JPS-PF, JPS-PF with static
%threshold criteria, and JLPS-PF
%very well. Hence in the next subsection,
%we only consider the analytical results.

\subsection{Comparison between the Proposed Scheme and Other Schemes}
 The
fraction of slot for probing one user is still set $\beta=0.1$. Here
four schemes are
 considered: (a) the proposed joint probing and scheduling scheme;
  (b) Round robin  scheduling; (c)
Genie-aided PF (GA-PF) scheme where full CQI is available to the
scheduler at the beginning of each slot; (d) Probe-all PF (PA-PF)
scheme where the transmitter probes all users before scheduling. For
both (c) and (d), the transmitter selects the user with the largest
$R_k(n)/T_k(n-1)$ for transmission. From \cite{borst_performance} we
know that the scheduling gain of GA-PF is  $ \mathbb{E} \left[
\max\limits_{k=1,\cdots,K} X_k \right]$. Then that of PA-PF is
$\max(1-K\beta, 0)\mathbb{E} \left[ \max\limits_{k=1,\cdots,K} X_k
\right]$.

%For JPS-PF, both the analytical results in Theorem \ref{theo_kappa}
%and the simulation results using scheme in Algorithm 1 are
%presented.

\begin{figure}[h]
\centering
\includegraphics[width=4.7in]{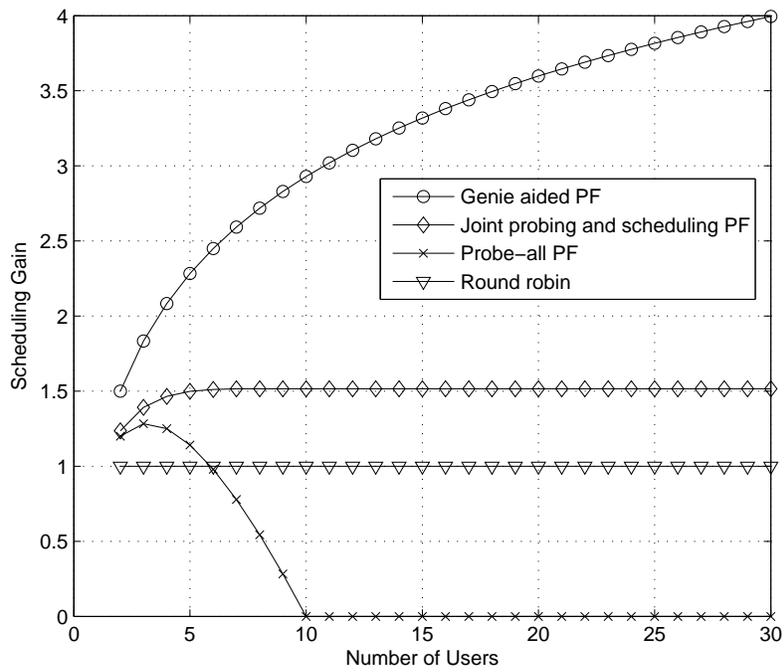}
\caption{Scheduling gain VS number of users.  $\beta=0.1$.}
\label{fig_gain}
\end{figure}

 Fig.   \ref{fig_gain}
presents the scheduling gain of schemes (a)-(d) as a function of the
number of users. We can  see from Fig.  \ref{fig_gain} that when
probing cost is taken into account, the scheduling gain does not
always increase but approaches to a limit value as the number of
users increases. This indicates that, by ignoring the cost of
channel probing, the ideal genie-aided PF does not reflect the
correct multiuser diversity characteristics. The comparison also
shows the advantage of the proposed joint probing and scheduling
scheme. For the probe-all PF scheme, it achieves higher gain than
round robin when the user population is not very large compared with
$\beta^{-1}$. However, when the number of user increases to some
extent, the scheduling gain of probe-all algorithm vanishes. That is
because almost all the period of one slot is used for user-probing
instead of data transmission.

\begin{figure}[h]
\centering
\includegraphics[width=4.7in]{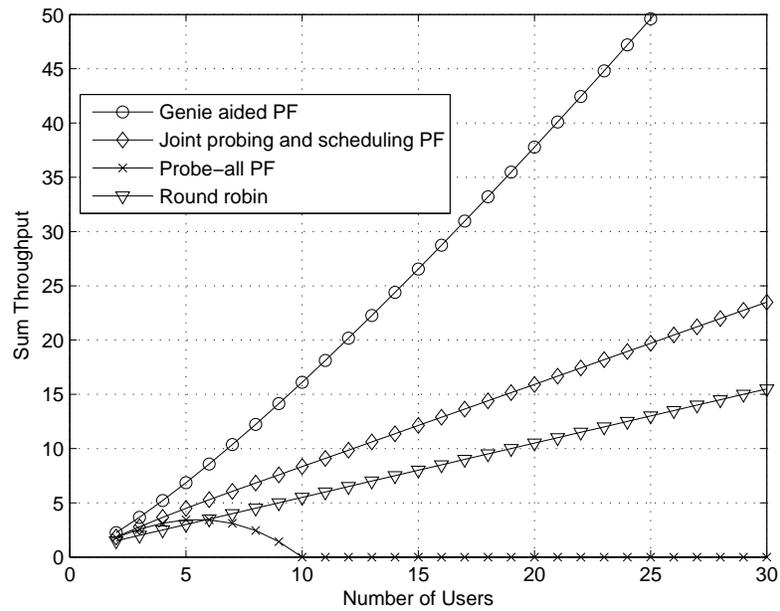}
\caption{Sum throughput VS number of users.   $\beta=0.1$.}
\label{fig_throughput} \vspace{-0pt}
\end{figure}

 Fig.  \ref{fig_throughput}  displays the sum throughput of
all schemes as the number of users increases. One can see that there
exists a relative large gap between the ideal genie-aided PF curve
and the proposed scheme. The gap quantifies the the extent to which
the user probing decreases the system performance. For example, when
the number of users is $K=20$, the throughput of the joint probing
and scheduling scheme only accounts for 55.64\% of that of the
genie-aided PF. And the throughput achieved by the joint scheme is
the highest among all the non-ideal schemes (a), (b) and (d). The
probe-all PF scheme performs similar to the joint probing and
scheduling scheme when there are not many users ($K \le 6$), but
degrades fast and even vanishes when the number of users becomes
large.

\section{Conclusion}
We have studied the problem of achieving proportional fairness in
wireless systems when explicitly taking into account the channel
probing cost. An optimal adaptive joint probing and scheduling
scheme is presented, as well as a static threshold based criteria
for determining whether to probe or to transmit. Using the
steady-state analysis, we have evaluated the scheduling gain
explicitly. Extension of the scheme to the case in which the
scheduler has no knowledge of the channel rate distribution has been
developed, which achieves almost the same performance of the
algorithm obtained under known rate statistics assumption and
outperforms other non-ideal PF schemes. In this work, we have
focused on the well-studied proportional fairness rule. It is
possible to extend the results to more general utilities, for
example, the $\alpha$ fair utility \cite{liu}. The  methodology
presented in this paper can then be carried through to that case as
well.

\appendices
\section{Proof of Theorem \ref{theo_monote}}
\label{app_proof_monote}
\begin{figure}[h]
\centering
\includegraphics[width=4.7in]{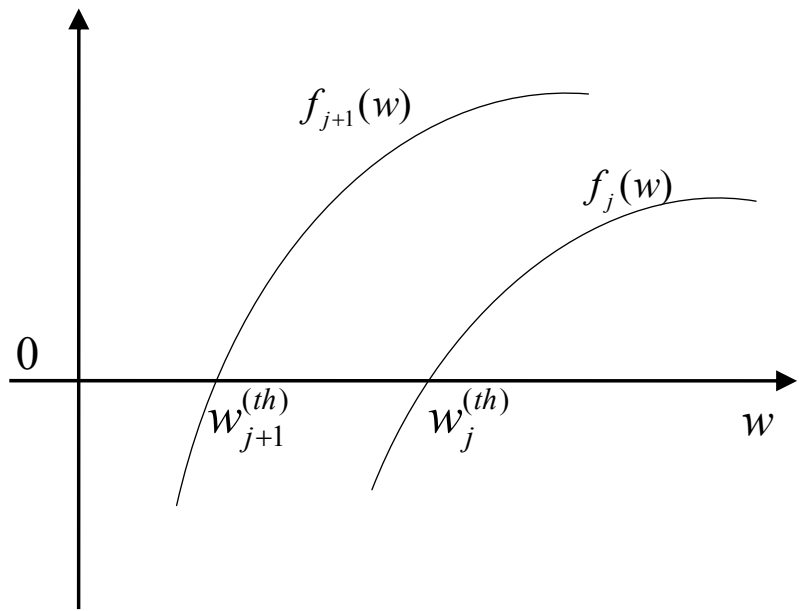}
\caption{Illustration of the property of function $f_j(w)$.}
\label{fig_fjw} \vspace{0pt}
\end{figure}

\begin{IEEEproof}
Let the largest throughput-normalized user rate after probing $j$
users be denoted by
\begin{align}
w_j = \max_{1\le l\le j} s_{k^{(l)}}
\end{align}
Then the current reward can be written as $y_j(s_{k_1},\cdots,
s_{k_j} ) = (1-j\beta)w_j$ and the expected reward obtained from
probing the next user is
\begin{equation}
\mathbb{E} [ y_{j+1}(s_{k_1},\cdots, s_{k_{j+1}} ) | s_{k_1},\cdots,
s_{k_j} ] = (1-(j+1)\beta) \mathbb E [ w_j \vee s_{k_{j+1}} | w_j ].
\end{equation}
 Then the event $\mathcal{E}_j$ can be expressed as
\begin{equation}
\mathcal{E}_j = \{ (1-j\beta)w_j \ge (1-(j+1)\beta) \mathbb E [ w_j
\vee s_{k_{j+1}} | w_j ] \}.
\end{equation}

We first show that there exists a threshold $w_j^{(th)}$ such that
the event $\mathcal{E}_j$ can be represented as $\mathcal{E}_j = \{
w_j \ge w_j^{(th)} \}$. To this end, let $f_j(w) = (1-j\beta)w -
(1-(j+1)\beta) \mathbb E [ w \vee s_{k_{j+1}}  ]$. Then $w \in
\mathcal{E}_j \Leftrightarrow f_j(w) \ge 0$. It is easy to verify
that $f_j(0)<0$ and $f_j(\infty)>0$. The function $f_j(w) $ can be
reorganized as $f_j(w) = \beta \mathbb E [ w \vee s_{k_{j+1}} ] +
(1-j\beta) \mathbb E [ w - w \vee s_{k_{j+1}} ]$. For any $w'>w>0$,
\begin{equation}  \nonumber
f_j(w')-f_j(w) = \beta \mathbb E [ w' \vee s_{k_{j+1}} -  w \vee
s_{k_{j+1}} ] + (1-j\beta) \mathbb E [ w'-w + w '\vee s_{k_{j+1}} -
w \vee s_{k_{j+1}}  ].
\end{equation}
Note that  $ w' \vee s_{k_{j+1}} \ge w \vee s_{k_{j+1}}$ and $w'-w
\ge w' \vee s_{k_{j+1}} - w \vee s_{k_{j+1}}$. Thus $f_j(w')-f_j(w)
\ge 0$, that is, $f_j(w)$ is a nondecreasing function. Summarizing
the properties of $f_j(w)$, it can be seen that the solution to
$f_j(w) \ge 0$ can be expressed as $w \ge w_j^{(th)}$.

We next show that $w_{j+1}^{(th)} \le w_{j}^{(th)}$. For  fixed $w$,
\begin{align}
& f_{j+1}(w) - f_j(w) \nonumber \\
 = &  (1-(j+1)\beta)w -
(1-(j+2)\beta) \mathbb E_{s_{k_{j+2}}} [ w \vee s_{k_{j+2}}  ] -
(1-j\beta)w + (1-(j+1)\beta) \mathbb E [ w \vee
s_{k_{j+1}}  ] \nonumber \\
 = & \beta \mathbb E_{s_{k_{j+2}}} [ w \vee s_{k_{j+2}} - w ]
   + (1-(j+1)\beta) \{ \mathbb E [ w \vee
s_{k_{j+1}}  ] - \mathbb E_{s_{k_{j+2}}} [ w \vee s_{k_{j+2}}
]  \} \nonumber \\
\ge & 0.
\end{align}
where the last `$\ge$' follows from the fact that $s_{k_{j+1}}$ and
$s_{k_{j+2}}$ are of the same type of distribution and $\mathbb E
s_{k_{j+1}} \ge \mathbb E s_{k_{j+2}} $. Note that $w_{j}^{(th)}$ is
the zero point of the function $f_j(w)$. Hence $w_{j+1}^{(th)} \le
w_{j}^{(th)}$, as illustrated in Fig. \ref{fig_fjw}.

Collecting the preceding results, we have $\mathcal{E}_j = \{ w_j
\ge w_j^{(th)} \} \subseteq  \{ w_{j+1} \ge w_j^{(th)} \} \subseteq
\{ w_{j+1} \ge w_{j+1}^{(th)} \} = \mathcal{E}_{j+1}$.
\end{IEEEproof}

%\section*{Acknowledgment}
%This work was partially supported by NSFC/RGC Joint Research Scheme
%No.60831160524.

%figure

%figure

%figure

%figure

%figure

%figure

%figure

%figure

% that's all folks
\end{document}